\documentstyle[12pt]{article}
\newlength{\dinwidth}
\newlength{\dinmargin}
\newlength{\extraspace}
\newlength{\extraspaces}
\newcommand{\be}{\begin{equation}
\addtolength{\abovedisplayskip}{\extraspaces}
\addtolength{\belowdisplayskip}{\extraspaces}
\addtolength{\abovedisplayshortskip}{\extraspace}
\addtolength{\belowdisplayshortskip}{\extraspace}}
\newcommand{\ee}{\end{equation}}
\newcommand{\bdm}{\begin{displaymath}
\addtolength{\abovedisplayskip}{\extraspaces}
\addtolength{\belowdisplayskip}{\extraspaces}
\addtolength{\abovedisplayshortskip}{\extraspace}
\addtolength{\belowdisplayshortskip}{\extraspace}}
\newcommand{\edm}{\end{displaymath}}
\renewcommand{\thefootnote}{\fnsymbol{footnote}}
\def\simlt{\mathrel{\lower2.5pt\vbox{\lineskip=0pt\baselineskip=0pt
           \hbox{$<$}\hbox{$\sim$}}}}
\def\simgt{\mathrel{\lower2.5pt\vbox{\lineskip=0pt\baselineskip=0pt
           \hbox{$>$}\hbox{$\sim$}}}}
\catcode`@=11
\newcount\@tempcntc
\def\@citex[#1]#2{\if@filesw\immediate\write\@auxout{\string\citation{#2}}\fi
  \@tempcnta\z@\@tempcntb\m@ne\def\@citea{}\@cite{\@for\@citeb:=#2\do
    {\@ifundefined
       {b@\@citeb}{\@citeo\@tempcntb\m@ne\@citea\def\@citea{,}{\bf ?}\@warning
       {Citation `\@citeb' on page \thepage \space undefined}}%
    {\setbox\z@\hbox{\global\@tempcntc0\csname b@\@citeb\endcsname\relax}%
     \ifnum\@tempcntc=\z@ \@citeo\@tempcntb\m@ne
       \@citea\def\@citea{,}\hbox{\csname b@\@citeb\endcsname}%
     \else
      \advance\@tempcntb\@ne
      \ifnum\@tempcntb=\@tempcntc
      \else\advance\@tempcntb\m@ne\@citeo
      \@tempcnta\@tempcntc\@tempcntb\@tempcntc\fi\fi}}\@citeo}{#1}}
\def\@citeo{\ifnum\@tempcnta>\@tempcntb\else\@citea\def\@citea{,}%
  \ifnum\@tempcnta=\@tempcntb\the\@tempcnta\else
   {\advance\@tempcnta\@ne\ifnum\@tempcnta=\@tempcntb \else \def\@citea{--}\fi
    \advance\@tempcnta\m@ne\the\@tempcnta\@citea\the\@tempcntb}\fi\fi}
\catcode`@=12
\newcommand{\SM}{Standard Model}

\newcommand{\bsg}{b \rightarrow s \gamma }
\newcommand{\la}{\lambda}
\newcommand{\La}{\Lambda}

\newcommand{\cL}{{\cal L}}
\newcommand{\cO}{{\cal O}}
\newcommand{\pr}{Phys.\ Rev.\ }

\newcommand{\np}{Nucl.\ Phys.\ {\bf B}}
\newcommand{\pl}{Phys.\ Lett.\ {\bf B}}

\newcommand{\ptp}{Prog.\ Theor.\ Phys.\ }

\begin{document}
\begin{titlepage}
\begin{flushright}
BUHEP-95-20\\
hep-ph/9506305\\
\today \\
\end{flushright}
\vspace{24mm}
\begin{center}
\Large{{\bf Flavor-Changing Neutral Current Constraints in
    Topcolor-Assisted Technicolor}}
\end{center}
\vspace{5mm}
\begin{center}
Dimitris Kominis\footnote{e-mail address: kominis@budoe.bu.edu}
\\*[3.5mm]
{\normalsize\it Dept. of Physics, Boston University, 590 Commonwealth
Avenue,}\\
{\normalsize\it Boston, MA 02215}
\end{center}
\vspace{2cm}
\thispagestyle{empty}
\begin{abstract}
It is argued that the topcolor models recently proposed by Hill
\cite{hill} may face significant constraints from flavor-changing
neutral current processes (such as \mbox{$B-\bar{B}$} mixing)
unless the mixing
angles between down-type quarks are small. The flavor-changing
processes are mediated by scalar bound
states which are likely to be light as a result of the near critical
dynamics of the $b$-quark sector. The consequences of
the latter on the bottom quark mass are also briefly discussed.
\end{abstract}
\end{titlepage}
\newpage

\renewcommand{\thefootnote}{\arabic{footnote}}
\setcounter{footnote}{0}
\setcounter{page}{2}

In topcolor models \cite{topc}, the top quark participates in a new strong
interaction which is spontaneously broken at some high energy scale
$\La$, but not confining. The strong dynamics leads to the formation of
a condensate $\langle \bar{t}_L t_R \rangle$ and gives rise to a large
dynamical mass for the top quark. If this top condensate is to be an
adequate source of electroweak symmetry breaking, then the scale $\La$
must be very high, of order $10^{15}$~GeV. This would render the
hierarchy $m_t/\La$ unnatural; that is, a certain coupling constant
would have to be tuned to an enormous precision for this hierarchy
to be achieved.

Recently, a class of models has been proposed \cite{hill} with the aim
of making the top quark mass generation natural.
The scale $\La$ was assumed to be of the order of only a TeV or
so -- in return, one had to give up the requirement that the
electroweak symmetry is broken solely by the top condensate.

In the models of Ref.~\cite{hill}, all quarks and leptons of the third
generation participate in the strong dynamics at scale $\La$.
Condensation occurs in the top sector (but not the bottom sector)
because of
isospin-violating $U(1)$ couplings which render the $b$-quark dynamics
subcritical. 
(Leptons are differentiated because they do not carry color.) Even so,
scalar bound states are expected to emerge at low energies,
coupling very strongly to the bottom quark.
Generally, these states are assumed to be heavy, with masses of the order
of the scale $\La$, and their effects on low-energy
processes are neglected. The purpose of this note is to point out
that, because 
the dynamics of the bottom quark sector cannot be far from
critical, these states are likely to be light and that, if there is
mixing between the third and the first two generations, they can mediate
observable flavor-changing neutral current (FCNC) processes, notably
mixing in the neutral $B_d$-meson system.

In the following paragraphs we review very briefly the main features of
the topcolor models of Ref.~\cite{hill}. Starting from an
effective four-fermion Lagrangian, we derive
the low-energy scalar bound state spectrum and calculate its
contribution to $B-\bar{B}$ mixing and to the partial width of the decay
$\bsg$. We also comment on the possible enhancement of
the $b$-quark mass due to near-critical interactions.

The models of \cite{hill} postulate an $SU(3)_1 \times SU(3)_2 \times U(1)_1
\times U(1)_2 \times SU(2)_L$ gauge structure at scales higher than a
few TeV. The third generation of quarks and leptons couples to $SU(3)_1
\times U(1)_1$ with the same quantum numbers as under ordinary QCD and
hypercharge, while the first two generations couple similarly to
$SU(3)_2 \times U(1)_2$. At a scale of order $\sim 1$~TeV,
the $SU(3)_1 \times
U(1)_1$ interactions are strong but not confining. An unspecified
mechanism that may or may not be relevant to electroweak symmetry
breaking drives the spontaneous breakdown of $SU(3)_1 \times SU(3)_2
\times U(1)_1
\times U(1)_2$ down to $SU(3)_{QCD} \times U(1)_Y$ at this scale. As a
result, a set of massive gauge bosons appears, a color octet $B^a$ and a
singlet $Z'$. At scales below the mass of these particles one can write
down an effective current-current interaction\footnote{We omit third
generation lepton currents, since we do not address the question of
lepton bound states in this paper.}
\be
\cL = -2\pi \left [ \frac{\kappa}{M^2_B} \left (\bar{t}\gamma_{\mu}
\frac{\la ^A}{2} t + \bar{b}\gamma_{\mu}\frac{\la ^A}{2} b \right )^2 +
\frac{\kappa_1}{M^2_{Z'}} \left(\frac{1}{6}\bar{\psi}_L \gamma_{\mu} \psi_L +
\frac{2}{3} \bar{t}_R \gamma_{\mu} t_R - \frac{1}{3} \bar{b}_R
\gamma_{\mu} b_R \right)^2 \right ]
\label{curcur}
\ee
where $M_B, M_{Z'}$ are the masses of the color octet and heavy $U(1)$
gauge bosons respectively, $\kappa, \kappa_1$ are the ``fine-structure
constants'' of the strong $SU(3)$ and $U(1)$ and
$\bar{\psi} = (\bar{t} \, , \, \bar{b} )$. After expanding and
Fierz-rearranging the terms in (\ref{curcur}), we obtain (taking
$M_{Z'}=M_B$ for simplicity)
\be
\cL = \frac{4\pi}{M^2_B} \left [ \left (\kappa + \frac{2\kappa_1}{9N_c}
\right )
\bar{\psi}_L t_R \bar{t}_R \psi_L + \left(\kappa - \frac{\kappa_1}{9N_c}
\right )
\bar{\psi}_L b_R \bar{b}_R \psi_L \right ]
\label{njl}
\ee
where $N_c$ is the number of colors.
Here we have only retained the terms involving Lorentz-scalar,
color-singlet fermion bilinears. It is these interactions that will be
relevant to our discussion, because they give rise to light scalar bound
states at low energies. Colored and vector bound states may arise from
other interactions contained in (\ref{curcur}), e.g. terms of the form
$\bar{q}\frac{\la^A}{2}q\bar{q}\frac{\la^A}{2}q$ or $\bar{q}_L
\gamma_{\mu} q_L \bar{q}_L \gamma^{\mu} q_L$. However, they are expected
to be heavier and their effects will not be discussed here.
Note that the leading isospin-violating term is proportional to $1/N_c$.
We have explicitly dropped higher orders in $1/N_c$; our subsequent
calculations are performed in the leading-$1/N_c$ approximation, even
though we will take $N_c=3$ in numerical estimates.

The Lagrangian (\ref{njl}) is equivalent \cite{eguchi} to one written
in terms of auxiliary complex doublet scalar fields $\phi_1$ and
$\phi_2$,
\be
\cL = \la_1 \bar{\psi}_L \phi_1 t_R + \la_2 \bar{\psi}_L \phi_2 b_R +
{\rm h.c.} - M^2_B (\phi_1^{\dag}\phi_1 + \phi_2^{\dag} \phi_2)
\label{effscalar}
\ee
where
\be
\la_1^2 = 4\pi (\kappa + \frac{2\kappa_1}{9N_c}) \mbox{\hspace{0.7cm}};
\mbox{\hspace{0.7cm}} \la_2^2 = 4\pi (\kappa - \frac{\kappa_1}{9N_c}).
\label{lambdas}
\ee
(By integrating out the static fields $\phi_1$ and $\phi_2$, one
can recover
the Lagrangian (\ref{njl}).) Equation~(\ref{effscalar}) gives the
effective Lagrangian at scale $M_B$. At lower scales the fields $\phi_1,
\phi_2$ develop kinetic terms and become dynamical; the induced (gauge
invariant) kinetic terms at a scale $\mu$ are
\be
\cL_{kin} = \sum_{i=1}^2 z_i(\mu) |D_{\mu}\phi_i|^2
\label{kin}
\ee
where
\be
z_i(\mu) = \frac{N_c \la_i^2}{16 \pi^2} \ln\frac{M_B^2}{\mu^2}.
\label{zis}
\ee
To study the nature of the chiral phase transition, one can calculate the
effective potential for $\phi_1, \phi_2$:
\be
V_{eff} = M_B^2 (1-\frac{N_c\la_1^2}{8\pi^2})
\phi_1^{\dag} \phi_1 +
M_B^2 (1-\frac{N_c\la_2^2}{8\pi^2}) \phi_2^{\dag} \phi_2 +
\frac{N_c}{16\pi^2} {\rm tr} \left \{ (X^{\dag}X)^2 \left(\ln
\frac{M_B^2}{X^{\dag} X} + \frac{1}{2} \right) \right \}
\label{effp}
\ee
where $X$ is a $4\times 4$ matrix with columns equal to $\la_1 \phi_1$
and $\la_2 \phi_2$:
\be
X = \left ( \la_1 \phi_1 \;\; \la_2 \phi_2 \right ).
\ee
A mass for the top quark, but not the bottom, is induced if $\phi_1$
acquires a vacuum expectation value, $f_0$ say, but $\phi_2$ does not.
This occurs if
\be
1 - \frac{N_c \la_1^2}{8\pi^2} < 0 \mbox{\hspace{1cm}} ;
\mbox{\hspace{1cm}} 1 - \frac{N_c \la_2^2}{8\pi^2}
> 0
\label{crit}
\ee
or, equivalently (cf. Ref.~\cite{hill}),
\be
\kappa + \frac{2\kappa_1}{9N_c} > \frac{2\pi}{N_c} \mbox{\hspace{1cm}} ;
\mbox{\hspace{1cm}} \kappa -
\frac{\kappa_1}{9N_c} < \frac{2\pi}{N_c}.
\label{crith}
\ee
Note that even though these conditions can be satisfied without severe
fine-tuning, the numerical factors are such that the bottom sector can
not be very far from criticality, unless $\kappa_1$ becomes very large.
In this case, however, besides potential problems with Landau poles, the
chiral symmetries of the $\tau$ lepton will be spontaneously
broken if $\kappa_1$ exceeds a critical value of $2\pi$,
and a large $\tau$ mass will be generated \cite{hilljts}.

The corrrectly normalized field $\tilde{\phi}_1 \equiv z_1^{1/2} \phi_1$
has a vacuum expectation value $f_{\pi}$ where
\be
f_{\pi}^2 = z_1 f_0^2 = \frac{N_c}{16\pi^2} m_t^2 \ln
\frac{M_B^2}{m_t^2} .
\label{fpi}
\ee
As mentioned earlier, the novel feature of the topcolor models
introduced in Ref.~\cite{hill} is that $f_{\pi}$ is no longer required to be
equal to the weak scale; in other words, top condensation need not be
the primary mechanism of electroweak symmetry breaking. For example, if $m_t
= 175$~GeV and $M_B = 1.5$~TeV, one obtains, for $N_c=3$, $f_{\pi} \approx
50$~GeV. Some other mechanism (such as Technicolor \cite{techni})
is needed in order
to ensure that the model correctly accounts for the observed $W$ and $Z$
masses.

We can write the doublets $\phi_1, \phi_2$ in terms of components fields
\be
\phi_1 = z_1^{-1/2} \left( \begin{array}{c} f_{\pi}+\frac{1}{\sqrt{2}} (
h + i\tilde{\pi}^0) \\ \tilde{\pi}^- \end{array} \right)
\mbox{\hspace{1.5cm}} \phi_2 = z_2^{-1/2} \left( \begin{array}{c} H^+ \\
\frac{1}{\sqrt{2}} (H+iA^0) \end{array} \right ).
\label{comps}
\ee
The $\tilde{\pi}^a$, or ``top-pions'', are Goldstone bosons of the
spontaneous breakdown of the chiral symmetries associated with $\psi_L$
and $t_R$. Small explicit chiral symmetry breaking terms (such as
ETC-generated masses \cite{etc} for the top and bottom quarks
in a technicolor scenario) will give mass to the top-pions,
estimated in Ref.~\cite{hill} to be of the order of a few hundred
GeV. The remaining degress of freedom have
masses which can be computed from the effective potential:
\begin{eqnarray}
& & m^2_h  =  4 \, m_t^2 \\
& & m^2_{H,A^0} =  M_B^2 \left (1 - \frac{N_c \la_2^2}{8 \pi^2}\right)
z_2^{-1} \label{ma} \\
& & m^2_{H^{\pm}} =  \left [ M_B^2 \left(1 - \frac{N_c \la_2^2}{8 \pi^2}
\right) +
\frac{N_c}{8\pi ^2}\la_2^2 m_t^2 \ln \frac{M_B^2}{m_t^2} \right ]
z_2^{-1} \label{mch}
\end{eqnarray}
The calculation of the effective potential (\ref{effp})
is done in the large-$N_c$
limit with a sharp momentum cutoff of $M_B$ in the fermion loop. It is
not clear how much such a calculation should be trusted, but it is
evident that the closer the coupling $\la_2$ is to the critical point,
the lighter the neutral scalars $H, A^0$ will be. For example, if we take
$\kappa =\kappa_1, M_B = 1.5$~TeV and 3 colors, these formulae give
$m_{H,A} \approx 230$~GeV, $m_{H^{\pm}} \approx 340$~GeV. As in
eq.~(\ref{fpi}), we evaluated the wavefunction renormalization constant
$z_2$ at
a scale $\mu=m_t$. So, these figures do not correspond to ``pole''
masses, but rather represent the inverse propagator at zero momentum. It
is this parameter that will appear in the effective Lagrangian for
$B_d^0$ to $\bar{B}_d^0$ transitions.

The Lagrangian (\ref{effscalar}), written in terms of component fields,
contains a term
\be
\cL_b = \frac{m_t}{f_{\pi}\sqrt{2}} \bar{b}_L (H+iA^0) b_R + {\rm h.c.}
\label{lb}
\ee
Note that the coupling of the scalars $H, A^0$ to the $b$-quark is
strong, proportional to $m_t$; if there are mixings between the bottom
quark and the $s$ and $d$ quarks, then the term above may induce strong
FCNC effects. At low energies, exchange of the $H$ and $A^0$ bosons
gives rise to an operator
\be
\delta\cL_{eff} = \frac{m_t^2}{2f_{\pi}^2} \frac{2}{m_H^2} \bar{b}_L b_R
\bar{b}_R b_L
\label{bbbb}
\ee
Note that $H$ and $A^0$ are degenerate. This degeneracy can be traced to
the (anomalous) $U(1)$ symmetry $b_R \rightarrow e^{i\alpha}b_R, \phi_2
\rightarrow e^{-i\alpha} \phi_2$ and will only be lifted by instanton
effects. We shall denote by $D_L (D_R)$ the mixing matrix between left-
(right-) handed flavor states (such as those appearing in (\ref{bbbb}))
and mass eigenstates. So
\be
\delta\cL_{eff}=\frac{m_t^2}{f_{\pi}^2 m_H^2} \left|(D_{Lbb}^*\bar{b}_L +
D_{Lbs}^*\bar{s}_L + D_{Lbd}^*\bar{d}_L ) (D_{Rbb}b_R + D_{Rbs} s_R +
D_{Rbd} d_R) \right |^2
\label{mass}
\ee
where $b_L, b_R$ etc. now denote mass states.
This leads to mixings in the neutral kaon and $B$-meson systems. In the
latter, the mass difference induced
between $B_d^0$ and $\bar{B}_d^0$ is given by
\be
\frac{\Delta m_B}{m_B} = \frac{5}{12} \frac{m_t^2}{f_{\pi}^2m_H^2}
\delta B_B f_B^2 \eta
\label{dmoverm}
\ee
where $\delta \equiv |D_{Lbd}^* D_{Rbb} D_{Rbd}^* D_{Lbb} |$, $\eta$ is
a QCD correction factor and $B_B$ is
defined by the hadronic matrix element
\be
\langle B^0 | \bar{d}_L b_R \bar{d}_R b_L | \bar{B}^0 \rangle =
-\frac{5}{12} B_B f_B^2 m_B^2
\ee
Experimentally, $\Delta m_B / m_B \approx 6.4 \times 10^{-14}$ \cite{pdb}.
Taking into account the Standard Model contribution to $\Delta m_B$
\cite{smmb}, an upper bound can be placed on $\delta/m_H^2$. For
$f_{\pi} = 50$~GeV, $(B_B f_B^2)^{1/2} = 180$~MeV \cite{bb}
and $\eta = {\cal O}(1)$,
the bound is
\be
\frac{\delta}{m_H^2} \simlt 1.3 \times 10^{-12} \;{\rm GeV}^{-2}
\label{bound}
\ee
While this has no consequences on the elements of the Kobayashi-Maskawa
matrix $K=U_L^{\dag}D_L$, a ``natural'' choice
$D_L \approx D_R \approx U_L \approx U_R \approx K^{1/2}$ would violate
the bound by approximately two
orders of magnitude, if $m_H$ is of the order of a few hundred GeV.
(In contrast, even this choice of mixing matrices is in reasonable
agreement with the observed $K-\bar{K}$ mass difference.) To
raise the masses of the bound states $H, A^0$, one would have to
increase the cutoff $M_B$, recovering some of the naturalness problems
the model sought to resolve at the outset, or else increase the $U(1)$
coupling $\kappa_1$. Since, as mentioned earlier, there is an upper
bound to the latter, it is unlikely that the mass $m_H$ can be pushed
into the TeV region.
Consequently, it will be difficult to accomodate, in the context of the
topcolor scheme under discussion,
models of fermion masses with large mixings in the
`down'-quark sector. A model where the off-diagonal elements of
the matrix $D_R$ are naturally suppressed, and thus the bound
(\ref{bound}) is respected, has been recently proposed by Lane and
Eichten \cite{lane}.
(It is of course possible that higher orders in
$1/N_c$ or gauge interactions shift the masses $m_H, m_{A^0}$
and alleviate some of the problems discussed.)

In Ref.~\cite{hill}, it was argued that the neutral top-pion $\tilde{\pi}^0$
may also couple to the $b$-quark through an instanton-induced term $\sim
\bar{b} \gamma^5 \tilde{\pi}^0 b$. One can readily obtain this term
using the same methods as above, but starting from an effective
Lagrangian that incorporates the effects of instantons \cite{hill}
\be
\cL \rightarrow \cL + \cL_{inst}\;\;,
\label{njl1}
\ee
\be
\cL_{inst} = \frac{k}{M_B^2} [\bar{b}_Lb_R \bar{t}_L t_R - \bar{t}_L b_R
\bar{b}_L t_R ] + {\rm h.c.}
\label{inst}
\ee
(We ignore an overall $CP$-violating phase.) The magnitude of the
coefficient $k$ is largely uncertain. Some QCD-based arguments indicate
that it is of order 0.1 or 1 \cite{hill}; however, we shall not base any
definitive conclusions on these estimates.
In terms of auxiliary fields $\phi_1, \phi_2$, the effective Lagrangian
(\ref{njl1}) is
\begin{eqnarray}
\cL &= &h_1 \,\cos\theta \, \bar{\psi}_L \phi_1 t_R + h_1 \,
\sin\theta
\, \bar{\psi}_L \phi_1^c b_R + h_2 \,\cos\theta \,\bar{\psi}_L
\phi_2 b_R \nonumber \\
 & & + h_2\,
\sin\theta \, \bar{\psi}_L \phi_2^c t_R + {\rm h.c.} -M_B^2
(\phi_1^{\dag}\phi_1 + \phi_2^{\dag} \phi_2 )
\label{fulll}
\end{eqnarray}
where
\begin{eqnarray}
\tan 2\theta & = & \frac{2k}{\la_1^2-\la_2^2} \\
h_1^2 & = & \frac{1}{2} [ \la_1^2 + \la_2^2 + (\la_1^2-\la_2^2) \sec
2\theta ] \\
h_2^2 & = & \frac{1}{2} [ \la_1^2 + \la_2^2 - (\la_1^2-\la_2^2) \sec
2\theta ]
\label{hth}
\end{eqnarray}
and $\la_1, \la_2$ are given in terms of $\kappa, \kappa_1$ in
eq.~(\ref{lambdas}). The effective potential, on the other hand, has the
form
\be
V_{eff} = M_B^2 (\phi_1^{\dag} \phi_1 + \phi_2^{\dag} \phi_2) -
\frac{N_c}{8\pi^2} M_B^2 \,{\rm tr} \,\Phi^{\dag} \Phi +
\frac{N_c}{16\pi^2} {\rm tr} \left \{ (\Phi^{\dag}\Phi)^2 \left(\ln
\frac{M_B^2}{\Phi^{\dag} \Phi} + \frac{1}{2} \right) \right \}
\label{effp1}
\ee
where $\Phi$  is a $4\times 4$ matrix with columns equal to
$h_1 \cos\theta \,\phi_1 + h_2 \sin\theta \,\phi_2^c$ and $h_2 \cos\theta
\,\phi_2 + h_1 \sin\theta \,\phi_1^c $.
Both $\phi_1$ and $\phi_2$ acquire vacuum expectation values.
Minimization of the potential yields gap equations for the
top and bottom quark masses (also easily derivable from the
Nambu--Jona-Lasinio Lagrangian (\ref{njl1})) which can be written
in the form
\begin{eqnarray}
m_t & = & \frac{k}{\la_2^2} \,m_b +
\frac{N_c}{8\pi^2}(\la_1^2-\frac{k^2}{\la_2^2}) \,m_t\,\left
(1-\frac{m_t^2}{M_B^2} \ln \frac{M_B^2}{m_t^2}\right) \label{gapt}\\
m_b & = & \frac{k}{\la_1^2} \,m_t +
\frac{N_c}{8\pi^2}(\la_2^2-\frac{k^2}{\la_1^2}) \,m_b\,
\left(1-\frac{m_b^{2}}{M_B^2} \ln \frac{M_B^2}{m_b^{2}} \right) .
\label{gapb}
\end{eqnarray}
Again, there is a phase where the top quark mass is large but the bottom
quark remains relatively light.
Provided $k$ is small compared to $\la_1$ and $\la_2$,
this phase is defined by conditions not very different from
the inequalities~(\ref{crit}).
Neglecting terms of order $m_b^{2}/M_B^2 \ln (M_B^2/m_b^{2})$,
and noting that $\la_1^2 \approx 8\pi^2 /N_c$, we obtain for $N_c=3$,
\be
m_b \approx \frac{k}{\la_1^2-\la_2^2}\, m_t = \frac{9k}{4\pi
\kappa_1}\,m_t  .
\label{bmass}
\ee
Note that this differs from the estimate given in \cite{hill} by a
factor $\la_1^2/(\la_1^2-\la_2^2)$;
nearly critical interactions enhance a fermion mass that originates in a
different sector of the dynamics (see also \cite{app}). An enhancement
by the same factor $\la_1^2/(\la_1^2-\la_2^2)$
occurs if an explicit $b$-quark mass is generated in the
theory (e.g. from ETC interactions).
Depending on the
magnitude of the coefficient $k$, the estimate in (\ref{bmass}) may or
may not represent too large a mass for the $b$-quark.

Since the top-pions are component fields of the linear combination
of $\phi_1$ and $\phi_2^c$ that gets all the vacuum expectation value,
it is evident that they couple to the (right-handed) $b$-quark like
\cite{hill}
\be
\frac{m_b^*}{f_{\pi}} (\frac{i}{\sqrt{2}} \bar{b}_L\tilde{\pi}^0 b_R +
\bar{t}_L \tilde{\pi}^+ b_R ) + {\rm h.c.}
\label{bpi}
\ee
The star on $m_b^*$ indicates that this is the piece of the $b$-quark
mass generated by instanton effects.
Comparison with (\ref{lb}) shows that the effects of the bound states
$H, A^0$ in $B-\bar{B}$ mixing will be by far more significant
than those of $\tilde{\pi}^0$ unless
$m_H/m_{\tilde{\pi}^0} \simeq m_t/m_b^*$ which seems unlikely, on the
basis of
the crude estimates reported here. 

As a final example of the effects of the bound doublet $\phi_2$, we
studied the process $\bsg$ in the context of the topcolor scenario of
Ref. \cite{hill}. In the absence of an instanton-induced $b$-quark mass,
the contribution of
top-pions to this decay has been calculated previously \cite{hill,balaji}.
The doublet $\phi_2$ couples strongly to the right-handed $b$-quark
and thus, after the appropriate rotation to mass eigenstates, gives rise
to an effective operator\footnote{The notation is based on that of Ref.
\cite{gsw}.}
\be
\cO_7^{\prime} = \frac{e}{16\pi ^2}m_b \bar{s}_R \sigma^{\mu \nu} b_L F_{\mu
\nu}
\label{osevenp}
\ee
which mediates the transition $\bsg$. In contrast, if $m_b^*=0$,
top-pion exchange induces the parity-conjugate operator $\cO _7 = (e/16
\pi^2) m_b \bar{s}_L \sigma^{\mu \nu} b_R F_{\mu \nu} $.
Both $\cO _7$ and $\cO_7^{\prime}$ are now
included in the effective Hamiltonian for
weak radiative $\bar{B}$ decays
\be
{\cal{H}}_{{\rm eff}} = \frac{4\,G_F}{\sqrt{2}}
V^*_{ts} \sum _j C_j(\mu) \cO _j  (\mu)
\label{heff}
\ee
The coefficient of the
operator $\cO_7^{\prime}$ in the effective Hamiltonian (\ref{heff}),
arising from the exchange of a virtual $H^+$ (see eq.(\ref{comps})),
turns out to be (ignoring any running between the scales
$m_{H^+}$ and $M_W$)
\be
C_7^{\prime}(M_W)= \frac{1}{6}\,\left ( \frac{v_w}{f_{\pi}} \right )^2
\frac{D_{Rbs}^*}{V_{ts}^*} A(x)
\label{c7p}
\ee
where $v_w=174$ GeV is the weak scale,
$x=m_t^2/m_{H^{\pm}}^2$ and the function $A(x)$ is given by \cite{smmb},
\be
A(x)=x \left [
\frac{\frac{2}{3}x^2+\frac{5}{12}x-\frac{7}{12}}{(x-1)^3} -
\frac{\frac{3}{2}x^2 -x}{(x-1)^4} \ln x \right ]
\label{hx}
\ee

Because of the different chirality structure of operators $\cO_7$ and
$\cO_7^{\prime}$, their effects add incoherently in the expression for the
partial width
\be
\Gamma (\bsg) = \frac{G_F^2 m_b^5 \alpha_{em}}{32 \pi^4} |V_{ts}|^2
\left (|C_7(m_b)|^2 + |C_7^{\prime}(m_b)|^2 \right ) .
\ee
Here $C_7$ is the coefficient of $\cO_7$ in the effective Hamiltonian
(\ref{heff}). If we assume that the QCD running of
$C_7$ and $C_7^{\prime}$ to the scale $m_b$ is of comparable magnitude
and we take $D_R = K^{1/2}$, then eq.(\ref{c7p}) represents an increase
of the \SM\ width by approximately 20\% for $m_{H^{\pm}}=350$ GeV
(falling to 7\% if $m_{H^{\pm}}=500$ GeV).

The presence of an instanton-induced interaction (eq.(\ref{inst}))
generates the coupling (\ref{bpi}) between the charged top-pion and the
right-handed $b$-quark, as well as a similar coupling between $H^+$
and $t_R$. As a consequence, there will be contributions to $C_7$ and
$C_7^{\prime}$
arising from both $\tilde{\pi}^+$ and
$H^+$ exchange. Evaluated at a scale $M_W$, these coefficients read
\begin{eqnarray}
C_7(M_W) &= &\frac{1}{2}A(x_W) - \left ( \frac{v_w}{f_{\pi}} \right )^2
\frac{D^*_{Lbs}}{V_{ts}^*} \left ( \frac{m_b^*}{m_b} (B(x_{\tilde{\pi}^+})-
B(x_{H^+})) -\frac{1}{6}A(x_{\tilde{\pi}^+}) \right ) \\
C_7^{\prime}(M_W) & = & \left ( \frac{v_w}{f_{\pi}} \right )^2
\frac{D_{Rbs}^*}{V_{ts}^*}
\left (\frac{1}{6}A(x_{H^+})+\frac{m_b^*}{m_b}(B(x_{H^+})-
B(x_{\tilde{\pi}^+}))
\right )
\label{c7c7}
\end{eqnarray}
where $x_i=m_t^2/m_i^2$ and the function $B(x)$ is
defined in \cite{gsw}:
\be
B(x)=\frac{x}{2} \left ( \frac{\frac{5}{6}x-\frac{1}{2}}{(x-1)^2} -
\frac{x-\frac{2}{3}}{(x-1)^3} \ln x \right ) .
\ee
We note a partial cancellation of the effects of the charged top-pion
and $H^+$. If, however, the $\tilde{\pi}^+$ is significantly lighter
than $H^+$, then potentially strong constraints on the mixing angles
$D_{Lbs}$ and $D_{Rbs}$ may arise. Note that, unlike eq.(\ref{bound}),
such constraints will not be imposed on the product of mixing angles and
therefore may not be evaded in models which naturally predict small
mixings in either the left- or right-handed `down'-quark sector.


In this paper we argued that the topcolor models of Ref.~\cite{hill}
may face constraints arising from FCNC processes, in particular
$B-\bar{B}$ mixing, mediated by the exchange of
scalar bound states coupling strongly to the $b$-quark. These
bound states are likely to be light, with masses of the order of a few
hundred GeV, unless the bottom quark dynamics is sufficiently
removed from criticality. This could also be desirable in order to avoid
excessive enhancement of the (instanton- or ETC-induced) $b$-quark mass.

I would like to thank B. Balaji, S. Chivukula,
C.~Hill and K.~Lane for valuable comments.
This work was supported in part under NSF contract
PHY-9057173 and DOE contract DE-FG02-91ER40676.

\end{document}